\documentclass[letterpaper, 10 pt, conference]{ieeeconf}  
\IEEEoverridecommandlockouts                              
\usepackage{amsmath,graphicx}
\usepackage{flushend}
\usepackage{epsfig}
\usepackage{flushend}
\usepackage{multirow}
\usepackage{color}
\usepackage{enumerate}
\usepackage{easylist}
\usepackage{amssymb }
\usepackage{amsmath}
\usepackage{bm}
\usepackage{cite}
\usepackage{float}
\usepackage{mathrsfs}
\usepackage{amsfonts}
\usepackage{graphicx}
\usepackage{subfigure}
\usepackage{algorithmicx}
\usepackage[table]{xcolor}
\usepackage[ruled,vlined]{algorithm2e}

\def\y{\bm{y}}
\def\X{\bm{X}}

\def\Z{\bm{Z}}

\def\C{\bm{C}}
\def\S{\bm{S}}
\def\c{\bm{c}}

\title{\LARGE \bf Siamese Neural Networks for EEG-based Brain-computer Interfaces}

\author{Soroosh Shahtalebi$^{1}$, Amir Asif$^{2}$, and Arash Mohammadi$^{1}$
\thanks{*This work was partially supported by the Natural Sciences and Engineering Research Council (NSERC) of Canada through the NSERC Discovery Grant RGPIN-2016-049988.}
\thanks{$^{1}$ S. Shahtalebi and A. Mohammadi are with the Concordia Institute for Information Systems Engineering (CIISE),  Concordia University, Montreal, QC, H3G-2W1, Canada. (\tt\small s\_shahta@encs.concordia.ca; arash.mohammadi@concordia.ca)}
\thanks{$^{2}$ A. Asif is with Department of Electrical and Computer Engineering, Concordia University, Montreal, QC, H3G-2W1, Canada (\tt\small amir.asif@concordia.ca).}
}
\begin{document}
\maketitle
\thispagestyle{empty}
\pagestyle{empty}
\begin{abstract}
Motivated by the inconceivable capability of human brain in simultaneously processing multi-modal signals and its real-time feedback to the outer world events, there has been a surge of interest in establishing a communication bridge between the human brain and a computer, which are referred to as Brain-computer Interfaces (BCI). To this aim, monitoring the electrical activity of brain through Electroencephalogram (EEG) has emerged as the prime choice for BCI systems. To discover the underlying and specific features of brain signals for different mental tasks, a considerable number of research works are developed based on statistical and data-driven techniques. However, a major bottleneck in development of practical and commercial BCI systems is their limited performance when the number of mental tasks for classification is increased. In this work, we propose a new EEG processing and feature extraction paradigm based on Siamese neural networks, which can be conveniently merged and scaled up for multi-class problems. The idea of Siamese networks is to train a double-input neural network based on a contrastive loss-function, which provides the capability of verifying if two input EEG trials are from the same class or not. In this work, a Siamese architecture, which is developed based on Convolutional Neural Networks (CNN) and provides a binary output on the similarity of two inputs, is combined with OVR and OVO techniques to scale up for multi-class problems. The efficacy of this architecture is evaluated on a $4$-class Motor Imagery (MI) dataset from BCI Competition IV$_{2a}$ and the results suggest a promising performance compared to its counterparts.
\end{abstract}

\section{Introduction}\label{sec:intro}
Unique capabilities of our brain in simultaneously processing different signal modalities in an efficient and real-time fashion have inspired a surge of interest to develop a direct communication medium between brain and outside world~\cite{makeig2012evolving, osuagwu2017implicit, arico2017passive}. Such a communication channel, referred to as Brain-Computer Interface (BCI), aims to monitor brain activities to extract hidden signatures embedded in recorded signals to be transformed into control commands/messages. The BCI is considered as a building block of human-in-the-loop cyber-physical systems~\cite{schirner2013future} with one ultimate goal of assisting impaired individuals to regain their lost functionalities. By capitalizing on the plasticity properties of brain~\cite{van2015brain}, BCI systems have opened up new perspectives in the field of rehabilitation/assistive technologies~\cite{daly2008brain,xu2014closed,grosse2011using,leeb2015towards} and have introduced/enhanced several new and existing rehabilitation therapies for neurophysiological disorders such as Autism Spectrum Disorder (ASD), Attention Deficit Hyperactivity Disorder (ADHD), Schizophrenia, and motor rehabilitation for post-stroke patients. Although promising in terms of potential impact on rehabilitation therapies~\cite{talakoub2017reconstruction}, BCI systems are still in their infancy and are incapable of providing the required performance for practical applications. This calls for an urgent quest to improve accuracy and bandwidth of BCIs.

Generally speaking, a BCI system consists of the following four main building modules: (i) Brain imaging; (ii) Pre-processing; (iii) Feature extraction, and; (iv) Feature translation. When it comes to selection of the imaging module (Item (i)), different modalities are available such as Electroencephalogram (EEG), Magnetoencephalogram (MEG), functional Magnetic Resonance Imaging (fMRI), and functional Near Infra-Red spectroscopy (fNIRs). Among these imaging modalities, EEG is considered as the prime choice for design of practical BCI systems due to its affordability, portability, and high temporal frequency.  For development of the remaining three modules, studying the motor activities of brain is of paramount importance. Motor tasks, either real or imagery (MI)~\cite{yao2017stimulus}, induce sensorimotor oscillations in the  motor cortex~\cite{edelman2016eeg}, referred to as Event Related Synchronization (ERS) and Event Related Desynchronization (ERD), which are detectable in EEG signals. ERD is a slow drop in the power of EEG signals, observed in $\mu$ band ($8-13$~Hz), and ERS is a fast rise of power observed in the $\beta$ band ($13-30$~Hz).

To develop BCI systems for rehabilitation purposes, accurate identification and extraction of ERD and ERS is crucially important, which explains development of several processing solutions in the literature, including Common Spatial Patterns (CSP)~\cite{ramoser2000optimal} and its extensions~\cite{ang2012filter,shahtalebi2018abayesian, lu2010regularized, lotte2011regularizing, aghaei2016separable, li2016unified, suk2013novel,shahtalebi2017ternary,shahtalebi2018bayesian,shahtalebi2019feature}. CSP technique derives a transformation matrix through a supervised process, which not only reduces the dimensionality of EEG signals but also minimizes the variation across each class and maximizes the distance between classes~\cite{xie2016motor}. CSP-based solutions are well suited for binary problems and although they generalize well on small datasets, their classification accuracy drops noticeably with the increase of the number of classes. More recently, successful application of artificial neural architectures in a variety of domains have ignited a surge of interest in utilization of Deep Learning (DL) methods for EEG processing applications. Despite the decent performance of existing DL-based methods in development of EEG-based BCI systems~\cite{schirrmeister2017deep}, the data-hungry nature of DL techniques limits their widespread and reliable application in practical settings.

To tackle the aforementioned issue, a branch in DL techniques is developed, referred to as Siamese Networks~\cite{bromley1994signature}. Siamese networks process two inputs in parallel and are specialized in detecting if the two inputs are drawn from the same class or not. The dual input strategy for Siamese networks drastically increases the number of training examples and enables us to take advantage of DL methods for small datasets. Stemmed in their successful application in verification and classification~\cite{hadsell2006dimensionality,lecun2005} tasks, a growing number of recent works in biomedical domain~\cite{tanaka2018signal,schirrmeister2017deep,patane2018calibrating,alaverdyan2020regularized,fares2019eeg,venturelli2017depth,stober2017learning,maiorana2019eeg} are dedicated to utilization of Siamese networks. In this paper and for the first time in the BCI domain,  to best of our knowledge, we propose to employ Siamese architectures for classification of Motor Imagery (MI) EEG signals. More specifically, our contributions are twofold: (i) Development of an algorithmic procedure to employ Siamese networks for multi-class classification problems, and; (ii) Successfully demonstrating feasibility of using Siamese networks in BCI applications, with the potential of enhancing the classification accuracy.

The rest of this paper is organized as follows. In Section~\ref{sec:methods}, the background and necessary concepts for our proposed method are discussed. Section~\ref{sec:arch}, presents the proposed  Siamese architecture developed for MI classification based on EEG signals. Simulation results are provided in  Section~\ref{sec:results}. Finally,  Section~\ref{sec:conc} concludes the paper.

\section{Problem Formulation}\label{sec:methods}
In this section, the technical foundations for development of our proposed Siamese architecture for EEG processing are briefly outlined. In Subsection~\ref{subsec:siamese}, basics of Siamese architectures along with the contrastive loss formulation are discussed. Subsection~\ref{subsec:multiclass}  briefly reviews a number of multi-class classification techniques. In what follows, scalars, vectors, and matrices are denoted by lowercase letters, bold lowercase letters, and bold uppercase letters, respectively.

\subsection{Siamese Networks}\label{subsec:siamese}
As discussed in Section~\ref{sec:intro}, in Siamese architectures, two identical neural networks with tied parameters are employed to process two input signals in parallel. The outputs are topped with an energy function, which measures the contrast between the two inputs. For the EEG classification problem at hand, we denote each EEG trial with $\{\X_k^{N_{ch} \times N_s}\}_{k=1}^{N_{trials}}$, where $N_{ch}$ and $N_s$ represent the number of channels and the number of samples, respectively. We model the preprocessing step as application of a nonlinear function $f(\cdot)$ on each trial $\X_k^{N_{ch} \times N_s}$ resulting in $\Z_k = f(\X_k)$ as the output of the pre-processing step.

To train and evaluate a Siamese network, in each iteration, we feed the network with $[\Z_1, \Z_2, Y]$, where $\Z_1$ and $\Z_2$ are two randomly selected EEG trials and $Y$ is the label denoting if they are from the same class or not. Modeling the effect of network as a function ($G_W(.)$) with parameters $W$, the Euclidean distance between the outputs of the network for the two inputs is calculated as follows
\begin{equation}
D_W(\Z_1,\Z_2) = ||G_W(\Z_1) - G_W(\Z_2)||_2.
\end{equation}
By taking $D_W$ as the short form of $D_W(\Z_1,\Z_2)$, the loss function for $p$ ($p \leq N^2_{trials}$) number of training pairs is defined as follows
\begin{equation}
l(W) = \sum_{i=1}^{p} L(W, (\Z_1,\Z_2,Y)^i),
\end{equation}
where $L(W,(\Z_1,\Z_2,Y)^i) = (1-Y)L_sD_W^i + YL_DD_W^i$. Terms $L_S$ and $L_D$ indicate the partial loss function for similar pairs and dissimilar pairs, respectively. The loss function is rewritten in the following form
\begin{equation}
L(W,\Z_1,\Z_2,Y) = (1-Y)\frac{1}{2}D_W^2 + (Y)\frac{1}{2}{max(0,m-D_W)}^2,
\end{equation}
where $m>0$ denotes a margin (radius around $G_W(\Z)$) to decide if a pair of signals are similar or not. In the above formulation, $m$ is a hyper-parameter of the network and $W$ is optimized through the training procedure. The number of training pairs $p$, could go up to the square of the number of available training samples. This completes a brief introduction to EEG signal processing via Siamese networks. Next, multi-class EEG classification problems are reviewed.

\subsection{Multi-class Classification Paradigms}\label{subsec:multiclass}
As discussed previously, the goal of the feature translation module in a BCI system is to correctly assign the extracted features in previous modules to physiological phenomena. Performance of classifiers normally degrades as the number of studied phenomena increases, which corroborates the urge for development of classifiers with higher learning capacities. Generally speaking, to handle multi-class classification problems, there are two main approaches: (i) Employment of classifiers that are naturally capable of handling multi-class problems, e.g., K-means, $K$ nearest neighbours (kNN), and Decision Trees to name but a few, and; (ii) Decomposing the multi-class problem into a number of binary classification problems, where binary classifiers could be employed. Due to the widespread utilization of binary feature extraction techniques in the BCI domain, e.g., CSP, the latter technique for multi-class problems is thoroughly investigated. To this aim, two techniques, i.e., One vs. One (OVO) and One vs. Rest (OVR), are typically used in the literature. In both methods, a coding matrix ($\C$) based on binary codewords is employed, which identifies the categorization of training trials into two supersets. Coding matrices for OVO and OVR are presented in Tables~\ref{table:ovo} and~\ref{table:ovr}, respectively. The number of columns in $\C$ determines the number of binary classifiers to be trained. For classifier $\Lambda^j$, two supersets $\S_0^j$ and $\S_1^j$ need to be formed, where all the classes denoted by $0$ and $1$ form supersets $\S_0^j$ and $\S_1^j$, respectively. Given the coding matrix $\C$ for a $4$-class problem and the set of training trials $\{\X_i, Y_i\}_{i=1}^p$, where $Y_i$ denotes the trial's label (numbers in the range $[1,2,3,4]$), supersets are formed as follows
\begin{equation}
\begin{cases}
\text{if}~~~ C^{y,j}~=~0:&~~~~~\X \rightarrow \S_0^j \\
\text{if}~~~ C^{y,j}~=~1:&~~~~~\X \rightarrow \S_1^j \\
\text{if}~~~ C^{y,j}~=~2:&~~~~~\text{No action.} \\
\end{cases}\label{eq:membership}
\end{equation}
In Tables~\ref{table:ovo} and~\ref{table:ovr}, the length of each codeword identifies the number of binary classification problems to be solved. As the proposed Siamese architecture provides a binary output indicating whether the two inputs are from the same superset or not, we believe that they are well-suited for the OVR and OVO approaches to solve a multi-class EEG classification problem. It is also worth mentioning that the classifier $\Lambda^j$ is in fact a Siamese network ($G_W^j(.)$), which is trained over the two supersets $\S_0^j$ and $\S_1^j$.
\begin{table}[]
\vspace{-0.1in}
\caption{OVO coding matrix for $4$ class problem.\label{table:ovo}}
\centering
\begin{tabular}{|c|c|c|c|c|c|c|}
\hline
Classes & C1 & C2 & C3 & C4 & C5 & C6 \\
\hline
Class 1 &1&1&1&2&2&2\\
Class 2 &0&2&2&1&1&2\\
Class 3 &2&0&2&0&2&1\\
Class 4 &2&2&0&2&0&0\\
\hline
\end{tabular}
\vspace{-0.1in}
\end{table}
\begin{table}[]
\caption{OVR coding matrix for $4$ class problem.\label{table:ovr}}
\centering
\begin{tabular}{|c|c|c|c|c|}
\hline
Classes & C1 & C2 & C3 & C4 \\
\hline
Class 1 &1&0&0&0\\
Class 2 &0&1&0&0\\
Class 3 &0&0&1&0\\
Class 4 &0&0&0&1\\
\hline
\end{tabular}
\vspace{-0.2in}
\end{table}

\section{Proposed Siamese Network for EEG Classification}\label{sec:arch}
In this section, the proposed Siamese architecture for EEG classification along with its algorithmic workflow in the training and testing phases are presented.

\begin{figure}[t!]
\vspace{-.1in}
\centering
\includegraphics[width=\linewidth]{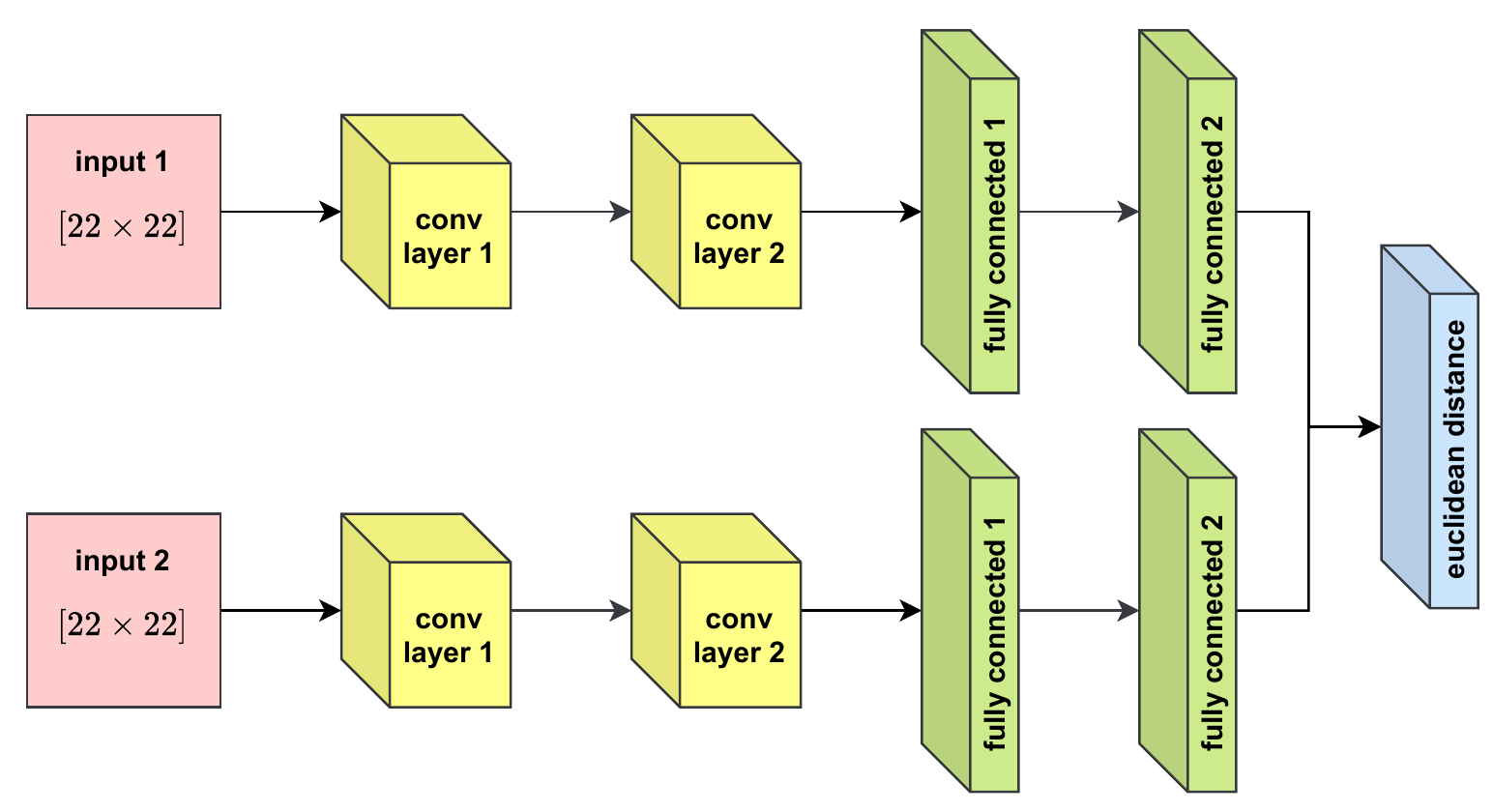}
\caption{The structure of the proposed Siamese network for EEG classification. The two parallel branches, within the network, share the same parameters and hyper-parameters. Input to the network is the covariance matrix of EEG signals. \textbf{Convolutional Layer 1:} [$16\times(3,3)$, stride=1, batch normalization, activation= Exponential Linear Unit (ELU)];  \textbf{Convolutional Layer 2:} [$32\times(3,3)$, stride=1, batch normalization, activation= ELU];  \textbf{Fully Connected Layer 1:} [$512$ units, activation= Rectified Linear Unit (ReLu), dropout=0.5], and; \textbf{Fully Connected Layer 2:} [$512$ units, activation= ReLu].}\label{fig:arch}
\vspace{-.2in}
\end{figure}

As stated previously, the main idea of the proposed architecture is to decompose the multi-class classification problem into a number of binary classification problems and then employ a Siamese architecture to design each binary classifier. In this work, the OVR and OVO techniques are employed for the decomposition task and a Siamese architecture is constructed based on a Convolutional Neural Network (CNN), which consists of two convolutional layers followed by two fully-connected layers. The network's architecture along with its hyper-parameters are shown in Fig.~\ref{fig:arch}. It should be noted that the hyper-parameters of the network are tuned through a rigorous parameter search with respect to classification accuracy over the validation set. It is worth noting that to validate the model in the development phase, a $5-$fold cross-validation approach is employed.

\begin{algorithm}[t]
\SetAlgoLined
\KwIn{Training EEG trials $\{\X_k^{N_{ch} \times N_s}\}_{k=1}^{N_{trials}}$}
\KwOut{Labels of unseen trials}
Obtain covariance of EEG trials\;
Select the coding matrix $\C$\;
\For{Each column of $\C$: $\c^j$}{
Split training trials to form supersets $\S_0^j$ and $\S_1^j$ according to Eq.~\eqref{eq:membership}\;
\For{Each trial in $\S_0^j$: $\Z_m$}{
\For{Each trial in $\S_1^j$: $\Z_n$}{
Determine $Y$ according to Eq.~\eqref{eq:Y}\;
Form $[\Z_m,\Z_n,Y]$
}
}
Train a Siamese network ($\Lambda^j~\text{or}~G_W^j$)\;}
\For{An unseen trial: $\Z_{test}$}{
\For{Each column of $\C$: $\c^j$}{
\For{Each trial in $\S_0^j$ and $\S_1^j$}{
Form $[\Z_{test},\Z_{{train}_i}]$\;
}
$y_{test}^j = \text{mode}(G_W^j([\Z_{test},\Z_{{train}}])) $\;}
$y_{test} = \underset{i}{\operatorname{argmin}}{\Delta(\y_{test},\c^i)}$  }

\caption{Siamese networks for BCI}\label{alg}
\end{algorithm}
\begin{table*}[t!]
\vspace{-0.1in}
\centering
\caption{Evaluation of the Siamese architecture for multi-class MI classification problem.}\label{table:res1}
\begin{tabular}{|c|c|c|c|c|c|c|c|}
\hline
Subject & Siamese OVR & Siamese OVO & SCSSP~\cite{aghaei2016separable} & FBCSP~\cite{aghaei2016separable} & FBCSP~\cite{ang2012filter} & BSSFO OVO~\cite{shahtalebi2018bayesian} & BSSFO OVA~\cite{shahtalebi2018bayesian} \\
\hline
Subject 1 &0.819&0.642&0.62&0.68&0.68&0.62& 0.31\\
Subject 2 &0.340&0.278&0.27&0.30&0.42&0.20& 0.08\\
Subject 3 &0.788&0.465&0.66&0.71&0.75&0.70& 0.57\\
Subject 4 &0.392&0.330&0.27&0.39&0.48&0.41& 0.45\\
Subject 5 &0.340&0.254&0.07&0.28&0.40&0.09& -0.07\\
Subject 6 &0.389&0.351&0.26&0.25&0.27&0.15& 0.08\\
Subject 7 &0.434&0.285&0.41&0.57&0.77&0.72& 0.62\\
Subject 8 &0.705&0.611&0.59&0.59&0.76&0.58& 0.30\\
Subject 9 &0.778&0.632&0.66&0.54&0.61&0.65& 0.55\\
\hline
Average &0.554&0.428&0.42&0.48&0.57&0.46&0.32 \\
\hline
\end{tabular}
\vspace{-0.1in}
\end{table*}
\begin{table}[t]
\vspace{-0.1in}
\centering
\caption{Evaluation results for binary classification of $4$ MI tasks.}\label{table:res2}
\begin{tabular}{|c|c|c|c|c|c|c|}
\hline
Subject &1 vs. 2&1 vs. 3&1 vs. 4&2 vs. 3&2 vs. 4&3 vs.4 \\
\hline
Subject 1 &79.44&91.00&91.71&93.60&100.0&59.86 \\
Subject 2 &63.88&65.78&64.80&62.92&62.56&63.99 \\
Subject 3 &88.20&83.00&89.77&86.12&92.84&64.64 \\
Subject 4 &53.93&65.55&76.47&74.41&70.56&54.11 \\
Subject 5 &54.94&56.26&58.51&61.53&54.49&53.81 \\
Subject 6 &65.67&61.56&65.64&65.69&64.61&62.43 \\
Subject 7 &51.69&78.46&78.29&74.83&73.37&56.57 \\
Subject 8 &91.81&78.58&93.15&76.89&81.25&73.86 \\
Subject 9 &82.76&94.30&97.24&76.06&85.75&84.31 \\
\hline
\end{tabular}
\vspace{-0.2in}
\end{table}

To train each binary classifier, $\Lambda^j$, supersets are formed according to the membership rule stated in Eq.~\eqref{eq:membership}. During the training phase, as we may encounter a case that the number of trials in $\S_0^j$ is not equal to the one for $\S_1^j$, we have introduced class weights to the loss function to compensate for the imbalanced data. To form the training package for each iteration, i.e., ($[\Z_1,\Z_2,Y]$), all of the possible cases to match a trial with another trial, either from the same superset or another, are collected. The label $Y$ is defined as follows
\begin{equation}
Y=
\begin{cases}
1~~~if~~~\Z_1,\Z_2 \in \S_0^j ~~or~~ \Z_1,\Z_2 \in \S_1^j \\
0~~~if~~~\text{otherwise.}\label{eq:Y}
\end{cases}
\end{equation}
In the evaluation phase, to identify the label of an unseen trial ($\Z_{test}$), the vote of each classifier ($\Lambda^j$) is collected separately and then the final label is constructed based on the collection of votes. For each classifier, all of the cases to pair the unseen trial with training trials are collected to form $\{[\Z_{test},\Z_{{train}_i}]\}_{i=1}^p$. Please note that in the OVO approach, the training trials of the classes, which are labeled by $``2''$, are not participated. All the collected pairs are fed to the trained Siamese network ($\Lambda^j$) and its binary output is collected to collectively form a binary vector of votes for the unseen trial. Based on the majority of votes, we conclude the final vote of each classifier for the test trial and finally, a codeword of length equal to the number of classifiers, denoted by $\y_{test}$,  is obtained. Finally, the label of the test trial is computed as
\begin{equation}
\underset{i}{\operatorname{argmin}}{~\Delta(\y_{test},\c^i)},
\end{equation}
where $\Delta$ and $\c^i$ denote the $L-1$ norm and the $i^{\text{th}}$ row of the coding matrix $\C$, respectively. A detailed algorithmic workflow of our proposed Siamese architecture for single-trial MI EEG classification tasks is outlined in Algorithm~\ref{alg}. This finalizes the workflow of our proposed Siamese architecture for single-trial MI EEG classification.

\section{Results}\label{sec:results}
In this section, our proposed Siamese architecture is evaluated through different scenarios and the details of implementation and the results are provided.  The employed dataset is taken from the BCI Competition IV (BCIC IV$_{2a}$) and consists of EEG recordings from $9$ healthy individuals. Four motor imagery tasks, including ``Right Hand MI'', ``Left Hand MI'', ``Foot MI'', and ``Tongue MI'' are investigated in this dataset resulting in a collection of $288$ trials for training and $288$ trials for testing. As part of the data collection procedure, a cue is shown on a screen to the participants to determine which MI task must be executed. The signals are recorded via $22$ EEG channels based on the $10-20$ electrode placement standard and the sampling rate of $250$ Hz. To prepare the data for our work, we dismissed the first $0.5$ seconds of signals after the cue onset to minimize the effect of activity in the visual cortex of brain on the studied motor phenomenon. Each EEG trial contains recordings from $0.5$ to $2.5$ seconds after the cue onset and forms a 2-dimensional matrix of size $22 \times 500$. A bandpass Butterworth filter of order $5$ in the frequency range of $7-30$ Hz is applied to the signals, and then, the covariance matrix of each trial is calculated as $\Z_i = (\X_i \X_i^T)/(\text{tr}(\X_i \X_i^T)$, where $\text{tr}(\cdot)$ denotes the trace of a matrix.

In the training phase, batches of $128$ pairs of trials are created and fed to the proposed Siamese network. To optimize the network, Adam optimizer is employed and the learning rate is set to $``0.0001''$. Our validation results show that $25$ epochs of training lead the network to an optimal balance between classification accuracy and the generalization over the studied phenomenon. It should be noted that based on our validations, the margin for similarity of two trials, $m$, is set to $0.5$. The proposed Siamese architecture is implemented on Keras~\cite{chollet2015keras} library in Python language.

The results of our evaluations on the BCIC IV$_{2a}$ dataset are presented in Tables~\ref{table:res1} and~\ref{table:res2}. Table~\ref{table:res1} provides a comparison between the performance of our proposed framework with renowned MI classification techniques, including FBCSP~\cite{ang2012filter,aghaei2016separable} and BSSFO~\cite{shahtalebi2018bayesian,suk2013novel}. In the FBCSP, the EEG signals are decomposed into $9$ different spectral bands and the frequency-specific features are obtained based on the CSP methodology. The BSSFO, on the other hand, employs a Bayesian optimization framework to derive subject-specific spectral filters to extract the most informative CSP-based features. In addition, the results in Table~\ref{table:res2} reflect the efficacy of our proposed deep learning-based Siamese architecture for binary classification of EEG signals. It should be noted that the results in Table~\ref{table:res2} represent the classification accuracy, whereas the results in Table~\ref{table:res1} show Kappa coefficient ($\kappa$) for classification accuracy. Kappa coefficient reveals the performance of a classifier compared to random labeling of unseen trials and is calculated as
\begin{eqnarray}
\kappa = (P_s - P_r)/(1 - P_r),
\end{eqnarray}
where $P_s$ is the probability of correct classification for the system and $P_r$ is the probability of random labeling of the unseen trials. It should be highlighted that in this work we have evaluated our proposed Siamese architecture in its most basic format to provide a proof-of-concept for this idea, thus several modifications and enhancements could be applied to the proposed technique to outperform the state-of-the-art results in this field, which will be the basis of our future works.

\section{Conclusions}\label{sec:conc}
In this work, we proposed a novel single-trial EEG classification framework based on a Siamese neural architecture. We constructed a Siamese network based on convolutional neural networks to provide a binary decision on whether two inputs are from the same class or not. Then, we employed the OVR and OVO techniques to tailor the Siamese networks for multi-class EEG classification problems. In this proof of concept, our results suggest a promising application for the Siamese architectures in the EEG classification tasks, which will be further investigated in future works.

\end{document}